\documentclass{article}
\usepackage{spconf,amsmath,amsfonts,mathtools,bbm}
\usepackage{graphicx}
\usepackage{spverbatim, bm}
\usepackage{amsmath}
\usepackage{amssymb}
\usepackage{booktabs}
\usepackage{xcolor, colortbl}
\usepackage{tabu}
\definecolor{Gray}{gray}{0.9}
\usepackage{comment}
\usepackage[hidelinks]{hyperref} 

\newcommand{\etal}{\textit{et al.}}

\title{Enhancing Image-Text Matching with Adaptive Feature Aggregation}
\name{Zuhui Wang \qquad Yunting Yin \qquad I.V. Ramakrishnan$^\dagger$\thanks{$\dagger$ Corresponding author}}
  
\address{State University of New York at Stony Brook, USA}

\begin{document}
%
\maketitle
\begin{abstract}
  Image-text matching aims to find matched cross-modal pairs accurately. While current methods often rely on projecting cross-modal features into a common embedding space, they frequently suffer from imbalanced feature representations across different modalities, leading to unreliable retrieval results. To address these limitations, we introduce a novel Feature Enhancement Module that adaptively aggregates single-modal features for more balanced and robust image-text retrieval. Additionally, we propose a new loss function that overcomes the shortcomings of original triplet ranking loss, thereby significantly improving retrieval performance. The proposed model has been evaluated on two public datasets and achieves competitive retrieval performance when compared with several state-of-the-art models. Implementation codes can be found \href{https://github.com/wzhings/itmAFA}{\underline{here}}. 
\end{abstract}


\section{Introduction}
\label{sec:intro}
Image-text matching which is also called image-text retrieval is a task to search for matched image-text pairs from a large-scale dataset. It is a fundamental task in the field of computer vision and natural language processing. It has a close relationship to many applications, such as image captioning~\cite{you2016image} and visual question answering~\cite{antol2015vqa}. The general idea of image-text matching is to extract features from different modalities and project these features into an embedding space to measure similarities between images and texts. Methods that are designed for image-text matching are usually evaluated on two retrieval tasks: text retrieval ({\it image-to-text}) and image retrieval ({\it text-to-image}).

\begin{figure}[t]
  \centering
  \includegraphics[width=\linewidth]{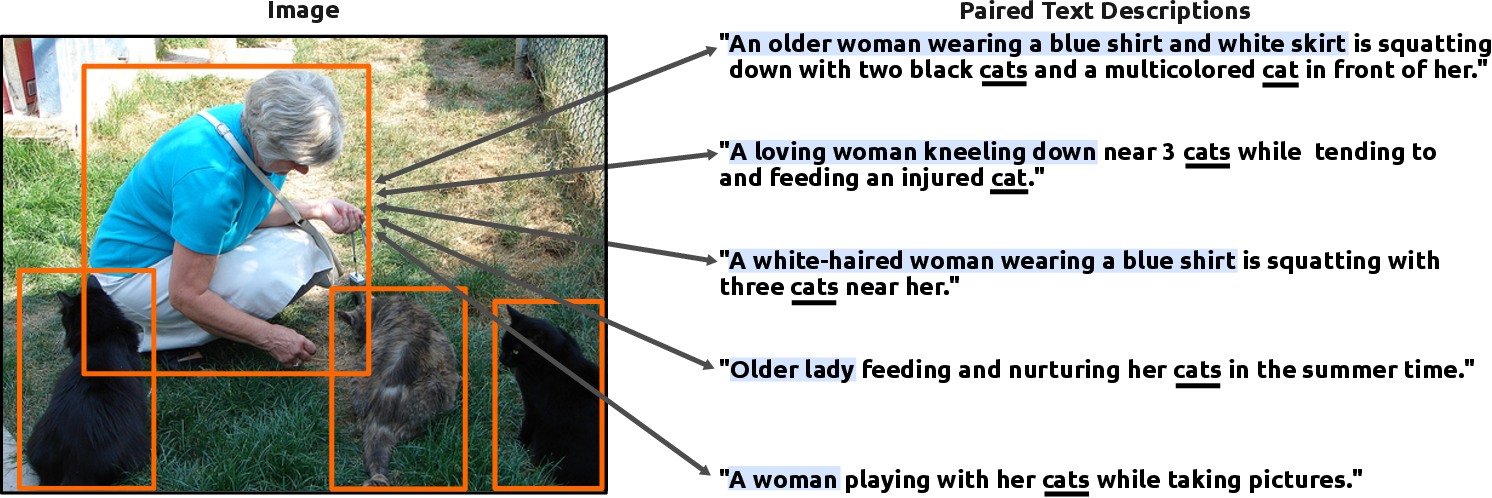}
   \caption{An example of positive image-text pairs. The same visual region can be described by different texts, and the same text (e.g., cat(s)) can be mapped to different visual regions.}
   \label{fig:f1}
\end{figure}

The main components of existing image-text matching models include feature extraction, feature enhancement, and similarity computation. Feature extraction is the first key module to obtain accurate retrieval results. Features from different modalities contain imbalanced information. First, images contain redundant information, but texts contain only necessary information. A given image can be described by multiple different texts. The paired texts normally contain partial information about the image. For example, as shown in Fig.~\ref{fig:f1}, the background lawn is not mentioned in any paired sentences. Second, for the same image, paired texts contain large variants of words and phrases. For example, the descriptions of the woman in Fig.~\ref{fig:f1} are different. Third, the feature extractors for images and texts have different representation abilities. The above reasons cause the model to have imbalanced feature representations, which is a key challenge to obtaining accurate retrieval results.  
 
Feature enhancement is the second key module to address the challenge of imbalanced representations. Most of the previous methods leverage complicated computation modules (e.g., cross-modal attentions~\cite{LeeCHHH18}) to improve feature representation abilities for cross-modal retrieval. However, these algorithms cannot solve the problem of imbalanced feature representations. They cannot aggregate single-modal features to obtain discriminative features for cross-modal retrieval. In this work, a new feature enhancement module is proposed to {\it adaptively} aggregate features for different modalities and obtain comprehensive features for similarity computations. 

\begin{figure*}[t]
  \centering
  \includegraphics[width=0.9\linewidth]{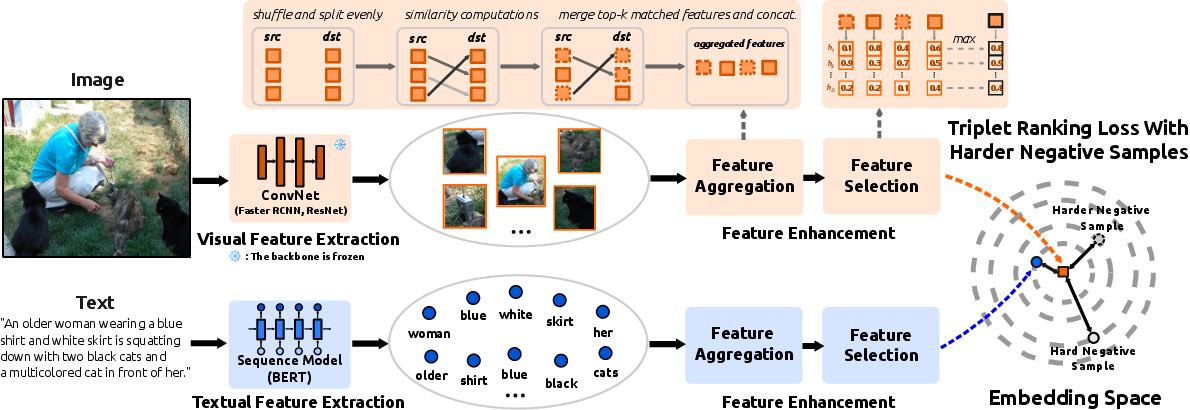}
  \caption{The proposed model architecture for the image-text matching task.}
   \label{fig:arch}
\end{figure*}

The second challenge of image-text retrieval comes from the loss function. Most of the previous works leverage the triplet ranking loss~\cite{faghri2018vse++} to separate positive image-text pairs from negative ones. The main limitation of this loss function is that only the {\it hard} negative samples of each {\it mini-batch} are considered. This loss design hinders the model from obtaining more discriminative features during inference. To address the above limitation, we propose a new loss function with {\it harder} negative samples to improve the model retrieval performance.

In this work, we propose a new image-text matching network to obtain accurate cross-modal retrieval performance. The main contributions are listed as follows: (1) A novel feature enhancement module is proposed to achieve accurate retrieval performance; (2) A new ranking loss function with harder negative samples is proposed to improve feature discriminative ability between positive and negative pairs; (3) The proposed network obtains competitive retrieval results on two public datasets.

\section{Related Work}
Most previous image-text matching works encode cross-modal features into a common embedding space for similarity computation~\cite{LiGFHX22, ChenH0JW21}. Feature extraction is the first key module to improve retrieval performance. Faghri~\etal~\cite{faghri2018vse++} proposed to extract visual features from hidden layers of Convolutional Neural Networks~\cite{HeZRS16} models and extract text information by  Recurrent Neural Networks (RNNs)~\cite{Alex2018}. Some works~\cite{LeeCHHH18, LiZLLF19, KarpathyL15} extend the network to include local visual information by extracting features based on detected regions of interest. 

Some previous works enhance feature representations with additional operations. It was proposed by \cite{JiCW21} to improve feature representations by cross-modal attention. Besides, Lee~\etal~\cite{LeeCHHH18} improved the cross-modal feature alignment by using stacked cross-modal attentions.  The matching scores were computed by both local and global alignments. Diao~\etal~\cite{DiaoZML21} leveraged graph reasoning to compute the similarity between cross-modal features. Most of these methods leverage complicated computation modules to improve feature representation abilities for cross-modal retrieval. 

Recently, some works proposed designing a new loss function to address the limitations of the triplet ranking loss. Li~\etal~\cite{Li2022} proposed to combine the triplet ranking loss and the vision-language contrastive learning loss~\cite{Oord2018} to obtain a unified loss function. Chen~\etal~\cite{Chen2022} proposed an intra-modal constraint network that reduces the distances of inter-modal positive pairs and increases the distances of intra-modal negative pairs. These methods only report performance improvement on several selected networks, and their effectiveness cannot be demonstrated by other existing methods that leverage traditional triplet ranking loss.

\section{Methodology}
\label{sec:method}


\label{subsec:feaExt}
\textbf{Feature extraction.} As shown in Fig.~\ref{fig:arch}, the model input consists of a pair of visual and textual information. Visual features are usually extracted by convolutional neural networks (e.g., ResNet~\cite{HeZRS16}). In this work, the input image was first processed to obtain the local regions of interest (ROI), and then convolutional features of each ROI were selected as the visual features~\cite{faghri2018vse++}. Input texts are normally processed by sequence models (e.g., RNNs~\cite{Alex2018} or Transformers~\cite{VaswaniSPUJGKP17, DevlinCLT19}) to extract sequential features. These local multi-modal features are written as follows,  
\begin{equation}
      \bm{f}_{\mathrm{vis}} = \mathbf{ConvNet}(\bm{x}), \quad \bm{f}_{\mathrm{txt}} = \mathbf{SeqNet}({\bm t}),
\end{equation}
where ${\bm x}$ is an input image, and $\bm{t}$ is the input text. ${\bm f}_{\mathrm{vis}}$ and $\bm{f}_{\mathrm{txt}}$ are the local features of detected visual ROI regions and text information, respectively. 

\label{sbusec: feaEnh}
\textbf{Feature aggregation.} The local visual and textual features are fed into two separate MLP modules~\cite{7103337} to improve their feature representation abilities. The MLP module is defined as follows,
\begin{equation}
  \bm{f}'_{\mathrm{vis}} = \mathbf{MLP}(\bm{f}_{\mathrm{vis}}), \quad \bm{f}'_{\mathrm{txt}} = \mathbf{MLP}(\bm{f}_{\mathrm{txt}}),
\end{equation}
where $\bm{f}'_{\mathrm{vis}}$ and $\bm{f}'_{\mathrm{txt}}$ are the enhanced visual and textual features.
The enhanced features contain local and redundant features from each modality. To reduce the feature redundancy, we propose to leverage the feature aggregation module to obtain discriminative features from each modality. As illustrated at the top of Fig.~\ref{fig:arch}. There are six main steps for the feature aggregation module: (1) enhanced single modal feature (e.g., $\bm {f}'_{\mathrm{txt}}$) are firstly randomly shuffled; (2) shuffled features are separated evenly by a half size into two subsets (i.e., the source ($\mathrm{src}$) subset, and destination ($\mathrm{dst}$) subset); (3) for each feature (e.g., here we use $\bm{f}^{\mathrm{src}}_a$ without loss of generality) in the $\mathrm{\it src}$ subset, compute the similarities to features in the $\mathrm{\it dst}$ subset. Then, select the sample (e.g., $\bm{f}^{\mathrm{dst}}_b$) from $\mathrm{\it dst}$ with the {\it highest} similarity for $\bm{f}^{\mathrm{src}}_a$ and get a matched pair ($\bm{f}^{\mathrm{src}}_a$, $\bm{f}^{\mathrm{dst}}_b$). Repeat such operations for each sample in the $\mathrm{\it src}$ subset; (4) sort all the matched pairs with similarities in descending order, and only keep the {\it Top-k} pairs with the highest similarity values; (5) merge the matched pairs by taking the average according to each dimension to create new features; (6) concatenate all the rest features together as new aggregated feature representations. Such an aggregation module can reduce feature redundancy (i.e., feature numbers) for each modality data, and obtain higher-level feature representations by aggregating local features.

\textbf{Feature selection.} After we have the aggregated features from each modality, we propose to select the most discriminative features from each dimension. The dimension-wise feature selection module is defined as follows, 
\begin{align}
  & h_i = \mathbf{MAX}_{\mathrm{dim}}(\bm{f}'_{\mathrm{vis/txt}}), \\
  & \bm{h}_{\mathrm{vis/txt}} = [h_1, h_2, \dots, h_D],
\end{align}
where $\bm{h}_{\mathrm{vis/txt}}$ is the final comprehensive representation for the visual or textual branch, $D$ is the dimension size of features, $h_i$ is the value for the $i$-th dimension, and $\mathbf{MAX}_{\mathrm{dim}}$ is the top dimensional-wise element selection operation.



\textbf{Triplet ranking loss with harder negative samples.} The triplet ranking loss~\cite{KarpathyL15} is primarily used to optimize the model by maintaining the distance between positive image-text pairs less than the distance between negative ones by a certain margin. \cite{faghri2018vse++} proposed to focus on the {\it hard} negative pairs during training, which is widely used for the cross-modal retrieval task. However, the main limitation of this loss function is the {\it hard} sample can only be selected within a mini-batch during training, which could easily cause the model to converge to the local minima, and affect testing retrieval performances. 
To overcome this limitation, we propose to construct some new samples (i.e., {\it harder} samples) based on existing samples in a mini-batch. Specifically, the {\it harder} samples are created by the mixup~\cite{ZhangCDL18} as follows,
\begin{align}
  \bm{h}^{\mathrm{g}}_{\mathrm{vis}} &= \lambda_1 \bm{h}_{\mathrm{vis}} + (1 - \lambda_1) \bm{h}_{\mathrm{txt}}, \\
  \bm{h}^{\mathrm{g}}_{\mathrm{txt}} &= \lambda_2 \bm{h}_{\mathrm{txt}} + (1 - \lambda_2) \bm{h}_{\mathrm{vis}}, 
\end{align}
where $(\bm{h}^{\mathrm{g}}_{\mathrm{vis}}, \ \bm{h}^{\mathrm{g}}_{\mathrm{txt}})$ is a pair of harder negative samples. $\lambda_i \sim \text{Beta}(t, t), \ i=1,2$. Additionally, $t$ is a hyperparameter of Beta distribution. The constructed harder samples can increase the chance of meeting more challenging negative samples than the real hardest negative samples in a mini-batch. The new loss function is defined by:
\begin{equation}
  \begin{aligned}
    \mathrm{Loss} &= \sum \left[\alpha_1 - \mathrm{s}(\bm{h}_{\mathrm{vis}}, \bm{h}_{\mathrm{txt}}) + \mathrm{s}(\bm{h}_{\mathrm{vis}}, \hat{\bm{h}}_{\mathrm {txt}}) \right]^+ \\
    &+ \left[\alpha_1 - \mathrm{s}(\bm{h}_{\mathrm{vis}}, \bm{h}_{\mathrm{txt}}) + \mathrm{s}(\hat{\bm{h}}_{\mathrm{vis}}, \bm{h}_{\mathrm{txt}}) \right]^+ \\
    &+ \left[\alpha_2 - \mathrm{s}(\bm{h}_{\mathrm{vis}}, \bm{h}_{\mathrm{txt}}) + \mathrm{s}(\bm{h}^{\mathrm{g}}_{\mathrm{vis}}, \hat{\bm{h}}^{\mathrm{g}}_{\mathrm{txt}}) \right]^+ \\
    &+ \left[\alpha_2 - \mathrm{s}(\bm{h}_{\mathrm{vis}}, \bm{h}_{\mathrm{txt}}) + \mathrm{s}(\hat{\bm{h}}^{\mathrm{g}}_{\mathrm{vis}}, \bm{h}^{\mathrm{g}}_{\mathrm{txt}}) \right]^+,
  \end{aligned}
\end{equation}
where $\alpha_1$ and $\alpha_2$ are two hyperparameters that stand for the margins between positive and negative pairs. We also define $\hat{\bm{h}}_{\mathrm{vis}} = \mathrm{argmax}_{{\bm{h}'}_\mathrm{vis} \neq \bm{h}_{\mathrm{vis}}} \mathrm{s}({\bm{h}'}_{\mathrm{vis}}, \bm{h}_{\mathrm{txt}})$ and $\mathrm{s}(\cdot, \cdot)$ is the cosine similarity. $\hat{\bm{h}}_{\mathrm{txt}} = \mathrm{argmax}_{{\bm{h}'}_\mathrm{txt} \neq \bm{h}_{\mathrm{txt}}} \mathrm{s}(\bm{h}_{\mathrm{vis}}, {\bm{h}'}_{\mathrm{txt}})$ as the hard negative visual and textual samples within a mini-batch. Similar definitions for harder negative examples (i.e., $\hat{\bm{h}}^{\mathrm{g}}_{\mathrm{vis}}$ and  $\hat{\bm{h}}^{\mathrm{g}}_{\mathrm{txt}}$) are also set.  $[x]^+ \equiv \mathrm{max}(0, x)$.

\section{Experiments}
\label{sec:exp}



\textbf{Datasets.} We evaluate the proposed approach on two public datasets, Flickr30K~\cite{Young2014} and MS-COCO~\cite{Lin2014}. Flickr30K contains 31,000 images that were collected from Flickr website. Each image is associated with five text descriptions. We follow the data split in~\cite{KarpathyL15}, where 29,000 images, 1,000 images and the rest 1,000 images are used for training, validation, and testing, respectively. MS-COCO contains 123,287 images, and each image is paired with five captions. We also use the data split of~\cite{KarpathyL15} where the dataset is split into 113,287 training images, 5000 validation images, and 5000 test images. For fair comparisons, we follow the result evaluation setting in~\cite{ChenH0JW21} where results of the MS-COCO dataset are reported in 1K and 5K. 1K results are the averaged results of five 1K folds. Additionally, we measure the cross-modal retrieval performance with a typical metric of Recall at K (\textsc{R@K}), where $K=\{1,5,10\}$. We also take the sum of all \textsc{R@K} of two retrieval tasks as \textsc{rSum}, which is often used to reflect the model's overall retrieval performances. 
 
\begin{table*}[htbp]
    \centering
    \fontsize{3}{3}\selectfont
    \caption{Retrieval results of state-of-the-art methods on Flickr30K and MS-COCO. The best and second best results (in \textsc{rSum}) are highlighted in \textbf{bold} and \underline{underline}.}
    \resizebox{\textwidth}{!}
    {
      \tabcolsep 2pt
      \begin{tabu} {@{\;} l @{\quad} ccccccc ccccccc@{\;}}
          \hline \\[-1.5ex]
  
          {\bf Datasets} & \multicolumn{7}{c}{\bf Flickr30K 1K Test} &
          \multicolumn{7}{c}{\bf MS-COCO 5-fold 1K Test} \\ 
          {Tasks} & \multicolumn{3}{c}{image-to-text} & \multicolumn{3}{c}{text-to-image} & & \multicolumn{3}{c}{image-to-text} & \multicolumn{3}{c}{text-to-image} \\
          {Methods} & \textsc{R@1} & \textsc{R@5} & \textsc{R@10} & \textsc{R@1} & \textsc{R@5} & \textsc{R@10} & {\textsc{rSum}} & \textsc{R@1} & \textsc{R@5} & \textsc{R@10} & \textsc{R@1} & \textsc{R@5} & \textsc{R@10} & \textsc{rSum} \\
          \hline \\[-1.5ex] 
           {VSE++~\cite{faghri2018vse++} \textsubscript{2018}} & {52.9} & {80.5} & {87.2} & {39.6} & {70.1} & {79.5} & {409.8} & {64.6} & {90.0} & {95.7} & {52.0} & {84.3} & {92.0} & {478.6} \\
           {SCAN~\cite{LeeCHHH18} \textsubscript{2018}} & {67.4} & {90.3} & {95.8} & {48.6} & {77.7} & {85.2} & {465.0} & {72.7} & {94.8} & {98.4} & {58.8} & {88.4} & {94.8} & {507.9} \\
           {VSRN~\cite{LiZLLF19}\textsubscript{2019}} & {71.3} & {90.6} & {96.0} & {54.7} & {81.8} & {88.2} & {482.6} & {76.2} & {94.8} & {98.2} & {62.8} & {89.7} & {95.1} & {516.8} \\
           {CVSE~\cite{WangZJPM20} \textsubscript{2020}} & {70.5} & {88.0} & {92.7} & {54.7} & {82.2} & {88.6} & {476.7} & {69.2} & {93.3} & {97.5} & {55.7} & {86.9} & {93.8} & {496.4} \\
           {SGRAF~\cite{DiaoZML21} \textsubscript{2021}} & {77.8} & {94.1} & {97.4} & \underline{58.5} & {83.0} & {88.8} & {499.6} & {79.6} & {96.2} & {98.5} & {63.2} & {90.7} & \textbf{96.1} & {524.3} \\ 
          GraDual~\cite{LongHWP22} \textsubscript{2022} & 76.1 & \underline{94.7} & \underline{97.7} & 57.7 & \underline{84.1} & \underline{90.5} & 500.8 & 76.8 & 95.9 & 98.3 & \underline{63.7} & 90.8 & 95.6 & 521.1 \\
          RCAR~\cite{DiaoZLRL23} \textsubscript{2023} & 77.8 & 93.6 & 96.9 & 57.2 & 82.8 & 88.5 & 496.8 & 78.2 & \underline{96.3} & 98.4 & 62.2 & 89.6 & 95.3 & 520.0 \\
          DivEmbed~\cite{KimKK23} \textsubscript{2023} & \underline{77.8} & 94.0 & 97.5 & 57.5 & 84.0 & 90.0 & \underline{500.8} & \underline{79.9} & 96.2 & \underline{98.6} & 63.6 & \underline{90.7} & 95.7 & \underline{524.6} \\ 
          \rowcolor{Gray}
          \textbf{Ours} & \textbf{81.8} & \textbf{95.8} & \textbf{98.1} & \textbf{59.9} & \textbf{84.9} & \textbf{91.2} & {\textbf{510.9}} & \textbf{80.1} & \textbf{96.3} & \textbf{98.7} & \textbf{64.0} & \textbf{90.9} & \underline{96.0} & \textbf{526.1} \\
          \hline
    \end{tabu}
  }
  \label{tab:t1}
  \end{table*}

\begin{table}[t]
    \centering
    \fontsize{8}{3}\selectfont
    \caption{Results on MS-COCO 5K. The best and second best results (in \textsc{rSum}) except ours are highlighted in \textbf{bold} and \underline{underline}. --: Results are missing from their publications. }
    \tabcolsep 3.5pt
    \begin{tabular}{@{\;}l@{\quad}ccccccc@{\;}}
         \toprule
         {\bf Datasets} & \multicolumn{7}{c}{\bf MS-COCO 5K Test}\\
         {Tasks}  & \multicolumn{3}{c}{image-to-text} & \multicolumn{3}{c}{text-to-image} \\
         {Method} & \textsc{R@1} & \textsc{R@5} & \textsc{R@10} & \textsc{R@1} & \textsc{R@5} & \textsc{R@10} & \textsc{rSum} \\
         \hline \\[-0.5ex] 
        VSE++ & 41.3 & 71.1 & 81.2 & 30.3 & 59.4 & 72.4 & 355.7 \\
         SCAN & 50.4 & 82.2 & 90.0 & 38.6 & 69.3 & 80.4 & 410.9 \\
         VSRN & 53.0 & 81.1 & 89.4 & 40.5 & 70.6 & 81.1 & 415.7 \\
         SGRAF & 57.8 & -- & \underline{91.6} & \textbf{41.9} & -- & 81.3 & -- \\
         RCAR & 57.4 & 83.8 & 91.0 & 40.7 & 69.8 & 80.4 & 423.1 \\
         DivEmbed & \underline{58.8} & \underline{84.9} & 91.5 & 41.1 & 72.0 & \textbf{82.4} & \underline{430.7} \\
         {{\textbf {Ours}}} & \textbf{59.3} & \textbf{85.9} & \textbf{92.3} & \underline{41.4} & \textbf{72.0} & \underline{82.3} & \textbf{433.2} \\   
         \bottomrule
    \end{tabular}
    \label{tab:t2}
\end{table}

\begin{table}[htbp]
  \centering
  \fontsize{8}{3}\selectfont
  \caption{Ablation studies on Flickr30K. {(I)}: Feature extractor and triplet ranking loss~\cite{faghri2018vse++}. {(II)}: Feature aggregation module; {(III)}: Feature selection module; {(IV)}: The new loss function.}
  {
    \tabcolsep 2pt
    \begin{tabu}{@{\;} cccc @{\quad} ccccccc @{\;}}
        \toprule
        \multicolumn{4}{l}{\bf Datasets} & \multicolumn{7}{c}{\bf Flickr30K 1K Test} \\ 
        \multicolumn{4}{l}{Tasks} & \multicolumn{3}{c}{image-to-text} & \multicolumn{3}{c}{text-to-image} \\
        (I) & (II) & (III) & (IV) & \textsc{R@1} & \textsc{R@5} & \textsc{R@10} & \textsc{R@1} & \textsc{R@5} & \textsc{R@10} & {\textsc{rSum}} \\
        \hline \\[-0.5ex]
        \rowcolor{Gray}          
        \checkmark & \checkmark & \checkmark & \checkmark & 81.8 & 95.8 & 98.1 & 59.9 & 84.9 & 91.2 & 510.9 \\
        \hline \\[-0.5ex]
        \checkmark & \checkmark & \checkmark & & 79.4 & 94.7 & 97.1 & 59.1 & 84.0 & 89.6 & 503.8  \\
        \checkmark & \checkmark & & & 75.3 & 92.1 & 95.7 & 57.1 & 82.7 & 89.4 & 492.4  \\
        \checkmark & & & & 70.3 & 91.0 & 95.5 & 51.9 & 79.0 & 87.2 & 474.9 \\   
        \bottomrule
  \end{tabu}
}
\label{tab:t3}
\end{table}

\textbf{Results and analysis.} Table~\ref{tab:t1} compares the proposed model with several state-of-the-art models on the Flickr30K and MS-COCO datasets for the text retrieval ({\it image-to-text}) and image retrieval ({\it text-to-image}) tasks. Based on the results, we can observe that our model achieves the highest total recall values (i.e., \textsc{rSum} values). For the Flickr30K dataset, the proposed model outperforms the second-best SOTA model (i.e., DivEmbed) by obtaining a large improvement from 500.8 to 510.9 on the overall \textsc{rSum}. Moreover, our method has higher values than the rest methods on all the individual recall metrics for the two retrieval tasks. 

In Table~\ref{tab:t1}, for the MS-COCO dataset, our method still obtains the highest \textsc{rSum} value compared with other methods. Specifically, our method has 1.5 (from 524.6 to 526.1) and 6.1 (from 520.0 to 526.1) higher values compared with DivEmbed and RCAR on \textsc{rSum}. Our method obtains higher values than other methods on the rest recall metrics for the two retrieval tasks only except 0.1 lower value on \textsc{R@10} for the image retrieval task when compared with the SGRAF model.

Moreover, Table~\ref{tab:t2} shows the performance of our proposed method on MS-COCO 5K test samples and illustrates our method has a better performance than the second-best model by a margin of 2.5 (from 430.7 to 433.2) in \textsc{rSum}. There exist small gaps (i.e., 0.5 for \textsc{R@1} and 0.1 for \textsc{R@10}) between our model and other models for the text retrieval task. However, our model has improvements on the rest recall metrics or the two retrieval tasks. The results demonstrate that our proposed model can obtain more competitive retrieval performances than other models on two retrieval tasks.

\begin{figure}[htbp]
  \centering
  \includegraphics[width=0.8\linewidth]{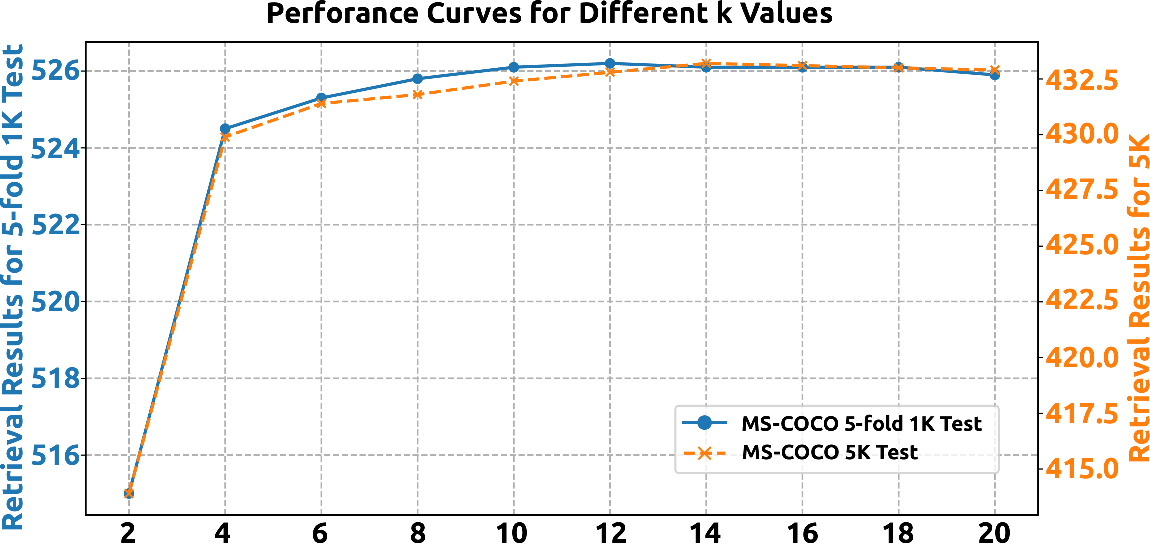}
  \vspace{-0.18in}
  \caption{Different k values for the Top-k aggregation in the proposed feature aggregation module.}
   \label{fig:topk}
\end{figure}

\textbf{Ablation study.} The experimental results are shown in Table~\ref{tab:t3}: (1) The first experiment replaces the proposed loss function with traditional triplet ranking loss, which reduces the retrieval values of 7.1 (from 510.9 to 503.8) on \textsc{rSum}. It demonstrates the proposed new loss function can improve the model performance on all the recall metrics. (2) The second replaces the proposed feature selection module by taking an average. The result shows that the performance of \textsc{rSum} drops by a margin of 11.2 (from 503.8 to 492.4). Based on the results of this experiment, we can observe that the feature selection helps the model select salient features for retrieval tasks. (3) The third experiment removes the proposed feature aggregation module. The \textsc{rSum} value dropped 17.5 (from 494.9 to 474.9). The result shows that the proposed feature aggregation strengthens feature representations and makes the aggregated features discriminative. Overall, the results in Table~\ref{tab:t3} demonstrate the feature enhancement module improves the model recall performance. The proposed new loss function can also further improve the model performance. We conducted additional experiments to investigate the effect of the Top-k aggregation in the feature aggregation module as shown in Fig.~\ref{fig:topk}. The results demonstrate that the Top-k aggregation can improve the model performance.

\section{Conclusion}
We propose a deep learning model for image-text matching and obtain competitive retrieval performances on two public datasets. The proposed model employs feature aggregation and selection to strengthen feature representations. Besides, a new triplet ranking loss with harder negative samples is proposed to improve the feature discriminative ability during model training. 
The proposed model will be extended to video-text retrieval in the future.

{\small
\bibliographystyle{IEEEbib}
\bibliography{strings,refs} 
}

\end{document}